# Radiative thermal switch via metamaterials made of vanadium dioxide-coated nanoparticles

Ming-Jian He[1,2], Xue Guo[1,2], Hong Qi[1,2,*], Lu Lu[3,**], and He-Ping Tan[1,2]

1 School of Energy Science and Engineering, Harbin Institute of Technology, Harbin 150001, P. R. China

2 Key Laboratory of Aerospace Thermophysics, Ministry of Industry and Information Technology, Harbin 150001, P. R. China

3 Hubei Key Laboratory for High-efficiency Utilization of Solar Energy and Operation Control of Energy Storage System, Hubei University of Technology, Wuhan 430068, China.

*Corresponding authors: Email: qihong@hit.edu.cn (H. Qi), lulu@hbut.edu.cn (L. Lu)

**Abstract:** In this work, a thermal switch is proposed based on the phase-change material vanadium dioxide ($VO_2$) within the framework of near-field radiative heat transfer (NFRHT). The radiative thermal switch consists of two metamaterials filled with core-shell nanoparticles, with the shell made of $VO_2$. Compared to traditional $VO_2$ slabs, the proposed switch exhibits a more than 2-times increase in the switching ratio, reaching as high as 90.29% with a 100 nm vacuum gap. The improved switching effect is attributed to the capability of the $VO_2$ shell to couple with the core, greatly enhancing heat transfer with the insulating $VO_2$, while blocking the motivation of the core in the metallic state of $VO_2$. As a result, this efficiently enlarges the difference in photonic characteristics between the insulating and metallic states of the structure, thereby improving the ability to rectify the NFRHT. The proposed switch opens pathways for active control of NFRHT and holds practical significance for developing thermal photon-based logic circuits.

The radiative heat flux can exceed the limit defined by Planck's law at the nanoscale due to the contribution of evanescent waves [1-3]. This phenomenon, known as the near-field radiative heat transfer (NFRHT), has garnered significant interest both in theoretical [4-6] and experimental studies [7-9]. NFRHT presents a unique heat transfer mechanism, holding promise for innovative energy conversion technologies and thermal management, such as near-field thermophotovoltaics [10], thermal modulation [11, 12] and more. Efficient methods for actively controlling NFRHT play a crucial role in micro/nanoscale thermal management [13, 14], holding substantial practical significance in enhancing the performance of functional devices [15, 16].

Nowadays, manipulating electric currents has become a common practice, while achieving precise and timely control of thermal flux remains challenging due to the fundamental difference between carriers for current and heat. Based on the unique mechanism of NFRHT, thermal analogues of the devices in electronics have been developed relying on the thermal photons, such as radiative thermal diodes [17, 18], thermal transistors [19] and others. Among these devices, a thermal analogue to electrical switch, capable of freely switching between low thermal resistance and high thermal resistance states, like On and Off modes of an electrical switch, is referred to as a radiative thermal switch. The underlying principle behind a radiative thermal switch is that NFRHT within the device can be alternated between ultra-low and ultra-high flux states, achievable through either active or passive methods. Active methods involve deliberate actions, such as twisting [20] or electrostatic gating [21]. While effective, these methods can be complex and highly reliant on precise operational control in practical applications [9]. On the other hand, passive methods exploit the unique characteristics of specific materials, such as nonreciprocal materials [22] or phase-change materials [23]. Passive approaches are comparatively easier to implement and do not require external actions for their operation, making them more practical and independent of precise manipulation.

Vanadium dioxide ($VO_2$), a widely-used phase-change material, has demonstrated its potential in developing innovative functional devices for NFRHT [17, 19, 23, 24]. These devices leverage unique property of existing in two distinct states of $VO_2$, below and above the phase-change temperature, allowing them to efficiently rectify NFRHT [25]. The performance of these devices primarily relies on the difference between these two states of $VO_2$. Nevertheless, there is no doubt that the performance of the functional devices is settled with the established difference between metallic and insulating $VO_2$. In this work, a novel $VO_2$ structure is proposed to enhance the ability to rectify the NFRHT by significantly enlarging the difference between the metallic and insulating states. This proposed structure functions as a radiative thermal switch and exhibits superior performance compared to traditional VO$_2$-based structures.

The schematic illustration of the proposed radiative thermal switch is presented in Fig. 1(a). The switch consists of two semi-infinite metamaterials containing nanoparticles (NPs) randomly dispersed within the host material. The nanoparticles, with a core-shell structure, are made of vanadium dioxide for the shell and a lossy metal core described by a Drude model, $\varepsilon(\omega) = 1 - \omega_p^2 / (\omega^2 + i\gamma\omega)$, where the plasma frequency is $\omega_p = 2.5 \times 10^{14}$ rad/s, and the damping coefficient is $\gamma = 10^{12}$ rad/s [26]. The cross section of the core-shell NPs is also given in Fig. 1(a), with the outside radius $R_{out}$ and the radius of the core $R_{in} = f_{cs} \times R_{out}$, where $f_{cs}$ denotes the filling factor of the core-shell NPs. $R_{out}$ is selected as 50 nm in the following simulations, unless otherwise specified. The electric and magnetic polarizabilities of the core-shell NPs can be expressed as follows [27]

$$\begin{aligned} \alpha_E &= 6\pi i a_1 / k_h^3 \\ \alpha_M &= 6\pi i b_1 / k_h^3 \end{aligned} \quad (1)$$

where $k_h$ is the wave vector in the host, $a_1$ and $b_1$ are the scattering electric and magnetic dipole terms. Specifically, for metamaterials made of VO$_2$ core-shell nanoparticles (VCS NPs), the axes of the anisotropic nanoparticles are randomly oriented in the host for the insulating VO$_2$. For this reason, the well-known 1/3 − 2/3 description is used to modify the polarizabilities of the VCS NPs [28]

$$\alpha_v = \frac{2}{3}\alpha_v(\varepsilon_\perp) + \frac{1}{3}\alpha_v(\varepsilon_\parallel) \quad (2)$$

where $v$ = E or M, $\varepsilon_\perp$ and $\varepsilon_\parallel$ are ordinary and extraordinary dielectric function component of uniaxial insulating VO$_2$, respectively [29]. The $2^m$-pole Mie coefficients of core-shell NPs are given by [27, 30]

$$\begin{aligned} a_m &= \frac{\psi_m(y)[\psi'_m(m_2 y) - A_m \chi'_m(m_2 y)] - m_2 \psi'_m(y)[\psi_m(m_2 y) - A_m \chi_m(m_2 y)]}{\xi_m(y)[\psi'_m(m_2 y) - A_m \chi'_m(m_2 y)] - m_2 \xi'_m(y)[\psi_m(m_2 y) - A_m \chi_m(m_2 y)]} \\ A_m &= \frac{m_2 \psi_m(m_2 x) \psi'_m(m_1 x) - m_1 \psi'_m(m_2 x) \psi_m(m_1 x)}{m_2 \chi_m(m_2 x) \psi'_m(m_1 x) - m_1 \chi'_m(m_2 x) \psi_m(m_1 x)} \\ b_m &= \frac{m_2 \psi_m(y)[\psi'_m(m_2 y) - B_m \chi'_m(m_2 y)] - \psi'_m(y)[\psi_m(m_2 y) - B_m \chi_m(m_2 y)]}{m_2 \xi_m(y)[\psi'_m(m_2 y) - B_m \chi'_m(m_2 y)] - \xi'_m(y)[\psi_m(m_2 y) - B_m \chi_m(m_2 y)]} \\ B_m &= \frac{m_2 \psi_m(m_1 x) \psi'_m(m_2 x) - m_1 \psi_m(m_2 x) \psi'_m(m_1 x)}{m_2 \chi'_m(m_2 x) \psi_m(m_1 x) - m_1 \psi'_m(m_1 x) \chi_m(m_2 x)} \end{aligned} \quad (3)$$

Here $\psi_m(z) = z j_m(z)$, $\xi_m(z) = z h_m(z)$, and $\chi_m(z) = -z y_m(z)$ relate the Riccati-Bessel functions to the spherical Bessel functions [27]. $m_1$ and $m_2$ represent the ratio of the refractive index of the core material and the shell material over that of the host material, respectively. The size parameters are defined as $x = k_h R_{in}$ and $y = k_h R_{out}$. The volume filling fraction $f$ is expressed as $f = 4\pi N R_{out}^3 / 3$, where $N$ is the volume density of the particles. The effective permittivity $\varepsilon^{eff}$ and permeability $\mu^{eff}$ of the metamaterials can be determined by relating them to the effective polarizabilities of the single nanoparticle using the Clausius-Mossotti model [27, 31],

$$\varepsilon^{eff} = \varepsilon_h \left(1 + N\left[\frac{k_h^3}{i6\pi a_1} - \frac{N}{3}\right]^{-1}\right)$$
$$\mu^{eff} = \mu_h \left(1 + N\left[\frac{k_h^3}{i6\pi b_1} - \frac{N}{3}\right]^{-1}\right)$$
(4)

where $\varepsilon_h$ and $\mu_h$ represent the permittivity and permeability of the host material, which are both taken as unity to avoid the influence of other factors.

The heat transfer coefficient (HTC) is used to quantitatively evaluate the NFRHT between the two metamaterials within the theory of fluctuation electrodynamics [1],

$$h = \frac{1}{4\pi^2} \int_0^\infty \hbar\omega \frac{\partial n}{\partial T} d\omega \int_0^\infty \xi(\omega,\kappa)\kappa d\kappa \tag{5}$$

where $\hbar$ is Planck's constant divided by $2\pi$, and $n = [\exp(\hbar\omega/k_B T)-1]^{-1}$ denotes the mean photon occupation number. $\xi(\omega, \kappa)$ is the energy transmission coefficient, in which the effective permittivity $\varepsilon^{eff}$ and permeability $\mu^{eff}$ defined with Eq. (4) take part in [31]. The vacuum separation distance between the two metamaterials is selected as $d = 100$ nm, considering that the effective medium theory can be applicable with $d > 2R_{out}/\pi$ [13].

VO$_2$ is a phase-change material known for its insulator-metal transition around 341 K. As depicted in Fig. 1(a), the thermal switch operates in the On mode when the VO$_2$ shell acts as an insulator below 341 K, and switches to the Off mode when the VO$_2$ shell behaves as a metal above 341 K. To demonstrate the thermal switching effect driven by the proposed structure, in Fig. 1(b), the spectral HTC for the On mode ($h_{on}$) and Off mode ($h_{off}$) are plotted with the colored-point and black-solid lines, respectively. Three different types of structures are considered here, (i) two slabs are both made of bulk VO$_2$, (ii) metamaterials composed of VO$_2$ NPs, and (iii) metamaterials made of VCS NPs. For VCS NPs, geometric parameters are optimized to $f_{cs} = 0.888$, and $f = 0.423$, which is consistent for case (ii). As can be seen from the black-solid lines, $h_{off}$ exhibits minimal values for all three structures. However, $h_{on}$ of VCS NPs is notably higher than the other two structures. It is evident that $h_{on}(\omega)$ exhibits three primary peaks near $0.85\times10^{14}$ rad/s, $1.2\times10^{14}$ rad/s, and $1.5\times10^{14}$ rad/s. To quantitatively evaluate the switching effect, a switching ratio defined as $\eta = (h_{on} - h_{off})/h_{on}$ is also presented in the figure. The switching ratios for structures (i)-(iii) are 43.37 %, 60.81 %, and 90.29 %, respectively. Remarkably, the VCS NPs achieve more than twice the switching ratio compared to the VO$_2$ slabs. This implies that the proposed structure made of VCS NPs exhibits significantly improved switching performance compared to traditional structures made of VO$_2$.

To gain a deeper understanding of the underlying mechanism responsible for the excellent switching performance, the energy transmission coefficients are depicted in Figs. 2(a)-(d) for various scenarios: slabs made of bulk VO$_2$, metamaterials filled with VCS NPs, VO$_2$ NPs, and NPs made of the Drude material used for the core, respectively. The left and right panels in Figs. 2(a)-(c) represent the results when VO$_2$ is in its metallic and insulating states, respectively. It can be seen from Figs. 2(a)-(c) that, the energy transmission coefficients are largest for the On mode VCS NPs. Corresponding to the three peaks of $h_{on}(\omega)$ observed in Fig. 1(b) for VCS NPs, the energy transmission coefficients also exhibit three branches near the three frequencies.

In Fig. 2(b), the most prominent $\xi(\omega, \kappa)$ branch near $1.5\times10^{14}$ rad/s can be considered as the evolution of the peak observed in Fig. 2(d), providing evidence that the insulating VO$_2$ shell effectively excites the surface plasmon modes induced by the core. The two branches near $0.85\times10^{14}$ rad/s and $1.2\times10^{14}$ rad/s in Fig. 2(b) can be viewed as the evolution and enhancement of the VO$_2$ modes in Figs. 2(a) or 2(c). This implies that the VO$_2$ modes can be excited significantly greater due to the coupling with the core. These results indicate that, for the insulating VCS NPs, the VO$_2$ shell substantially interacts with the core, resulting in a remarkable enhancement in thermal photon-based energy exchange. As shown in the left panels of Figs. 2(a) and 2(c), the VO$_2$ acts as a metal, leading to weak energy transmission coefficients that primarily remain in the low wave vector space. In Fig. 2(b), it is nearly the same with that of Fig. 2(c), indicating that the metallic VO$_2$ shell completely inhibits the surface plasmon modes of the core, thus blocking the energy exchange, and causing the device to function in its Off mode.

In Fig. 3(a), the switching ratios are calculated for different volume filling fraction of the NPs and filling factor of the core-shell NPs. The results demonstrate that as $f$ increases, the switching performance improves. However, the key factor influencing the switching performance is not $f$, but rather $f_{cs}$. As observed from the results, the switching ratios increase as $f_{cs}$ enlarges. However, the varying trend of the switching ratios $\eta$ with respect to $f_{cs}$ is not always monotonic, and there exists an optimal value for $f_{cs}$ near $f_{cs} = 0.9$. This suggests that only a thin VO$_2$ shell can lead to excellent switching performance. It can be concluded that one should select a structure with an optimized filling factor of the VCS NPs to achieve high switching performance. The optimized $f_{cs}$ represents a trade-off between the size of the VO$_2$ shell and the core, maximizing the difference in plasmonic characteristics between metallic and insulating VCS NPs. The interplay between the NPs and the plasmonic system is governed by the local density of electromagnetic states (LDOS). In Figs. 3(b) and 3(c), the LDOS near the NPs $\rho^E$, relative to the LDOS in free space $\rho_0^E$, is given for an isolate VCS NP and VO$_2$ NP, respectively. The ratio $\rho^E / \rho_0^E$ can be regarded as the enhancement of LDOS near the nanoparticle. The results indicate that the LDOS enhancement of the insulator is much greater than the metal for VCS NPs, compared with that of VO$_2$ NPs. This serves as clear evidence that the insulating VO$_2$ shell can effectively couple with the core to induce plasmonic excitation, thus significantly enlarging the difference between the metallic and insulating states of the structure compared to traditional VO$_2$ structure. Consequently, the ability to rectify NFRHT has been greatly improved, resulting in higher switching ratios.

To evaluate the performance of the proposed thermal switch under different conditions, the switching ratios are plotted as a function of the separation distance $d$ in Fig. 4 for various cases. It should be noted that the size of the NPs is adjusted when the vacuum separation distance $d$ is below 100 nm to satisfy the effective medium theory with $d > 2R_{out}/\pi$ [13]. The blue and red lines represent the results for VO$_2$ slabs and VCS NPs with optimized parameters ($f$ and $f_{cs}$). It is demonstrated that the switching effect of the proposed thermal switch is always better than that of VO$_2$ slabs in the near-field regime with $d <$ 500 nm. The dashed lines correspond to the switching ratios for different volume filling fraction of the NPs. In the strong near-field regime, there is almost no difference between them and the line with the optimized parameters. However, as $d$ increases, the switching ratios decrease with smaller $f$, consistent with the observations in Fig. 3(a). The other two lines denote the switching ratios when the permittivity of the host material is selected as $\varepsilon_h$ = 3 and $\varepsilon_h$ = 5. The results demonstrate that the switching effect of the proposed structure weakens when $\varepsilon_h >$ 1. Nonetheless, it remains higher than that of the VO$_2$ slabs in the near-field regime, providing potential for the practical application of the proposed switch.

In summary, the present work proposes a radiative thermal switch with a significantly improved switching ratio, more than 2 times higher compared to traditional VO$_2$ slabs, reaching 90.29% with a vacuum separation distance of 100 nm. The key factor responsible for this excellent switching effect lies in the role of the VO2 shell. It effectively couples with the core, resulting in the strong modes that enhance heat transfer at the insulating state, while blocking the surface plasmon modes of the core in the metallic state of VO$_2$. This mechanism greatly amplifies the difference in plasmonic characteristics between the insulating and metallic states of the structure, thus enhancing the ability to rectify the NFRHT. The proposed structure opens up potential applications in active control of NFRHT, particularly in thermal photon-driven logic circuits and thermal management.

## DATA AVAILABILITY

Data will be made available on request.

## ACKNOWLEDGEMENTS

The supports of this work by the National Natural Science Foundation of China (No. 52206082), China Postdoctoral Science Foundation (No. 2021TQ0086), the Natural Science Foundation of Heilongjiang Province (No. LH2022E063), Postdoctoral Science Foundation of Heilongjiang Province (No. LBH-Z21013), Excellent Thesis of Masters and Doctors of New Era Heilongjiang Province (No. LJYXL2022-009) are gratefully acknowledged.

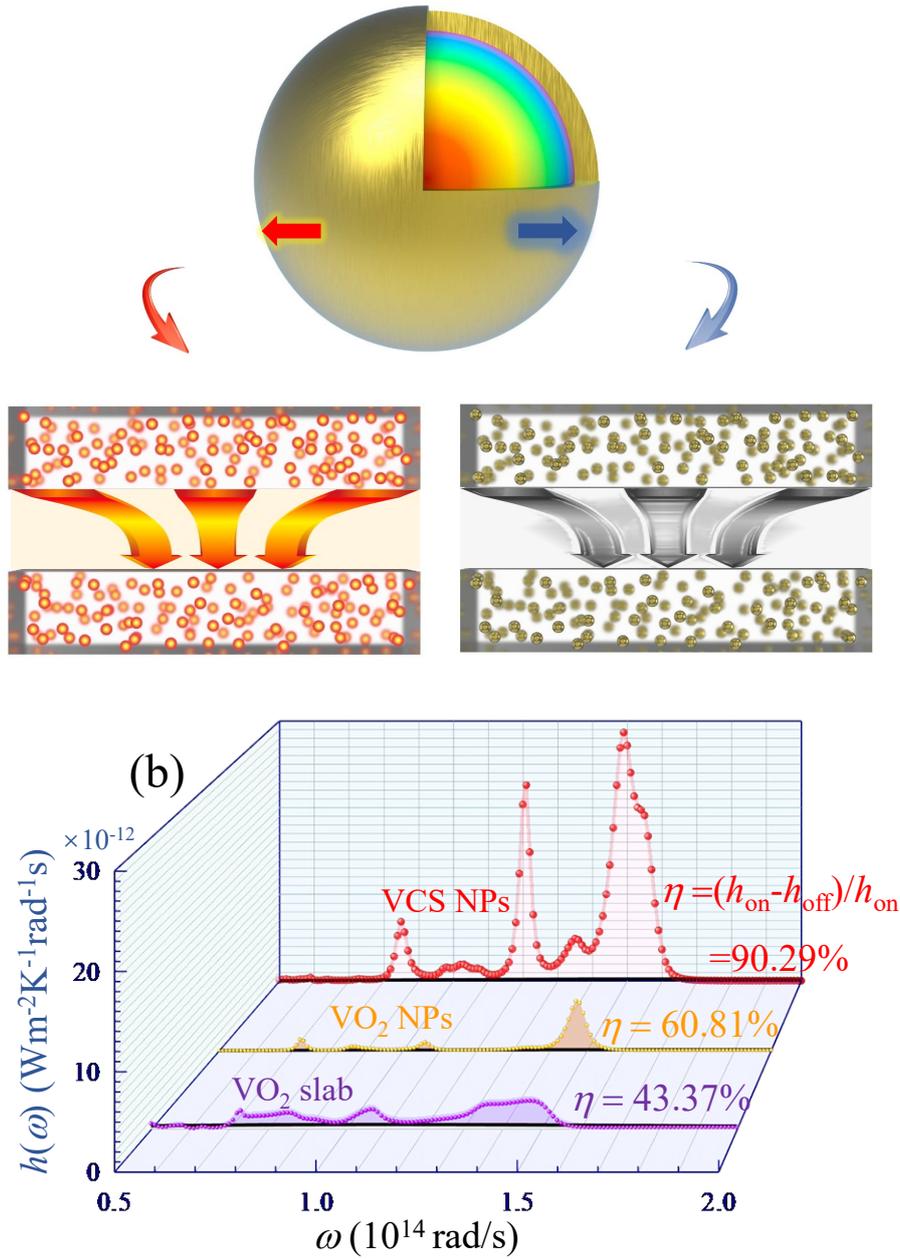

FIG. 1 (a) Schematic illustration of the proposed radiative thermal switch at On ($T < 341$ K) and Off ($T > 341$ K) modes, with the cross section of the $VO_2$-coated core-shell nanoparticle. (b) The spectral heat transfer coefficients for the thermal switch made of $VO_2$ slab, metamaterials filled with $VO_2$ NPs, and VCS NPs. The colored-point line and black-solid line denote the results for On and Off modes, respectively.

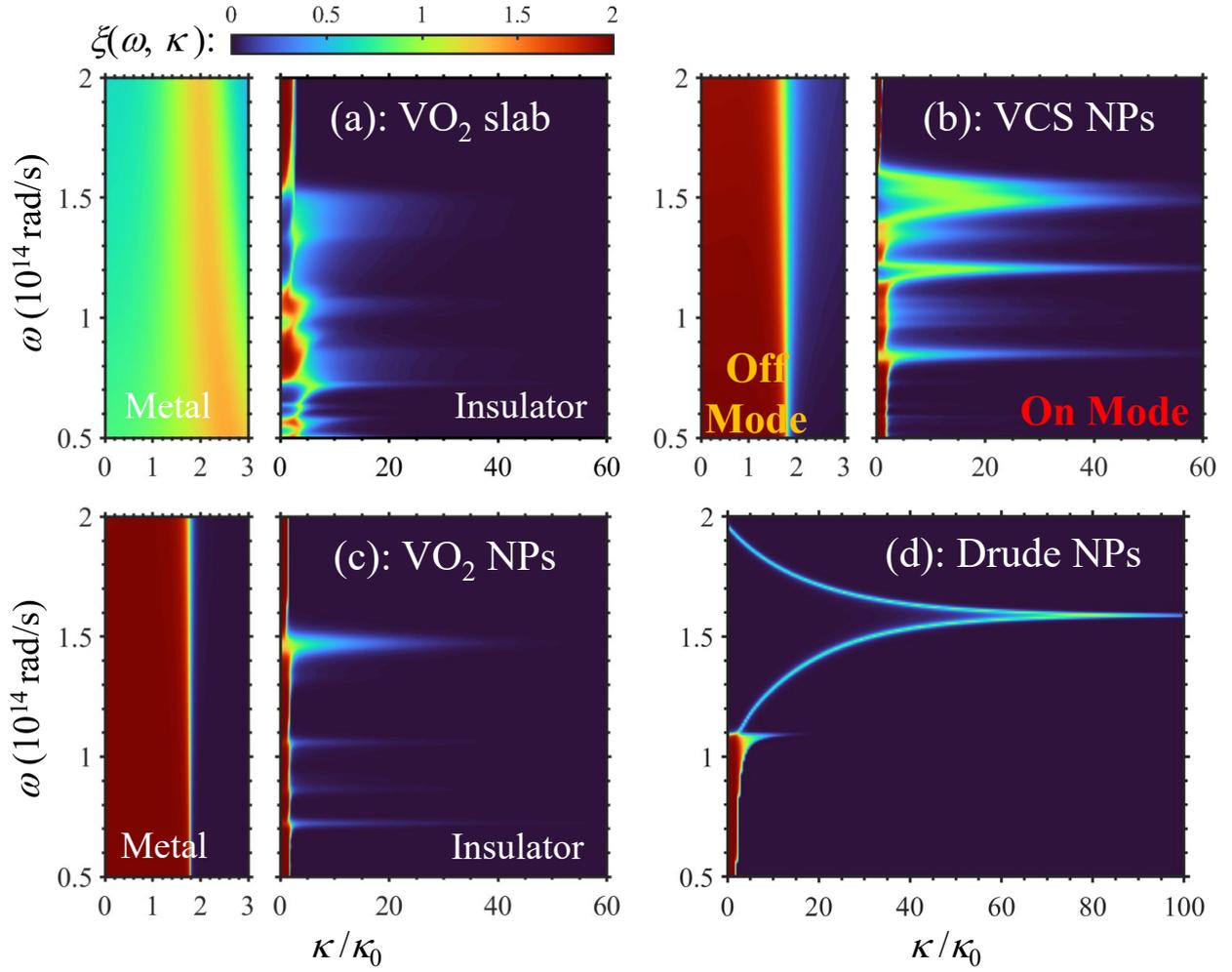

FIG. 2  The energy transmission coefficients for (a) VO$_2$ slab, (b) VCS NPs, (c) VO$_2$ NPs, and (d) Drude NPs, respectively.

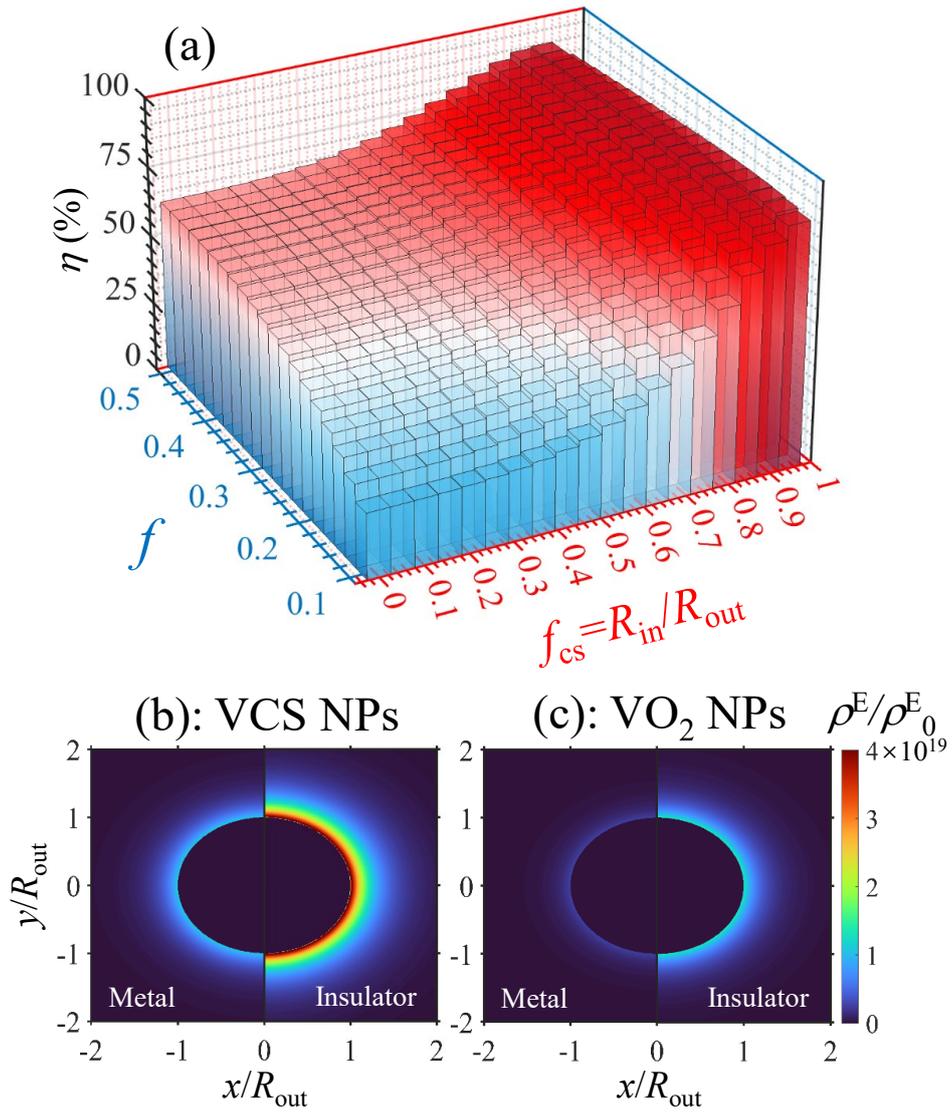

FIG. 3 (a) The switching ratios $\eta$ for different volume filling fraction $f$ and filling factor of the core-shell NPs $f_{cs}$. The local density of electromagnetic states enhancement near the isolated nanoparticles for (b) VCS NPs and (c) VO$_2$ NPs.

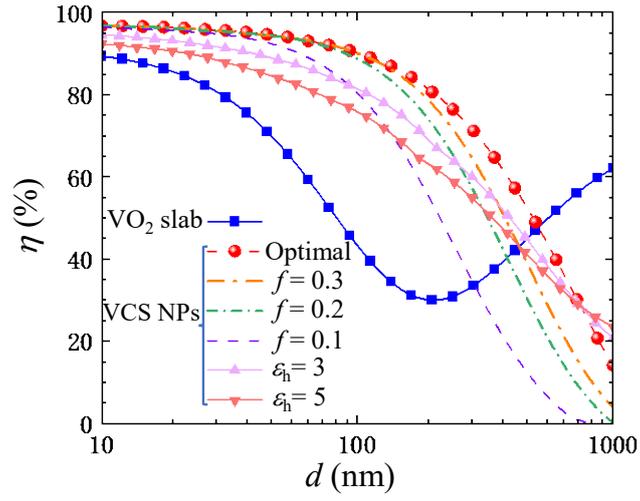

FIG. 4　The switching ratios $\eta$ at different vacuum separation distances $d$ for $VO_2$ slab and VCS NPs in different situations.